\documentclass[showpacs,preprint,preprintnumbers,showpacs,showkeys,superscriptaddress,amsmath,amssymb,nofootinbib]{revtex4-1}
\usepackage{amsmath}
\usepackage{latexsym}
\usepackage{color}
\usepackage{latexsym}
\usepackage{amsmath}
\usepackage{amssymb}
\usepackage{euscript}
\usepackage[latin1]{inputenc}
\usepackage{graphics}
\usepackage{graphicx}
\usepackage{picture}

\newcommand{\be}{\begin{equation}}
\newcommand{\ee}{\end{equation}}
\newcommand{\ba}{\begin{eqnarray}}
\newcommand{\ea}{\end{eqnarray}}

\def\uma{\rm 1\!\!\hskip 1 pt l}
\begin{document}


\title{\Large{Minimally Extended Left-Right Symmetric Model for Dark Matter with U(1) Portal}}

\author{M. J. Neves}\email{mariojr@ufrrj.br}
\affiliation{Departamento de F\'isica, Universidade Federal Rural do Rio de Janeiro,
BR 465-07, 23890-971, Serop\'edica, RJ, Brazil}
\author{J. A. Hela\"yel-Neto}\email{helayel@cbpf.br}
\affiliation{Centro Brasileiro de Pesquisas F\'isicas, Rua Dr. Xavier Sigaud
150, Urca,\\
Rio de Janeiro, Brazil, CEP 22290-180}
\author{Rabindra N. Mohapatra}\email{rmohapat@umd.edu}
\affiliation{Maryland Center for Fundamental Physics and Department of Physics\\
University of Maryland, College Park, MD 20742, USA}
\author{Nobuchika Okada}\email{okadan@ua.edu}
\affiliation{Department of Physics and Astronomy, University of Alabama, Tuscaloosa, AL 35487, USA}




\begin{abstract}
\noindent

A minimal extension of the left-right symmetric model for neutrino masses that includes a vector-like singlet fermion dark matter (DM)  is presented with the DM connected to the visible sector via a gauged $U(1)$ portal. We discuss the symmetry breaking in this model and calculate the mass and mixings of the extra heavy neutral gauge boson at the TeV scale. The extra gauge boson can decay to both standard model particles as well to dark matter. We calculate the relic density of the singlet fermion dark matter and its direct detection cross section and use these constraints to obtain the allowed parameter range for the new gauge coupling and the dark matter mass.
\end{abstract}


\maketitle

\pagestyle{myheadings}
\markright{Minimally Extended Left-Right Symmetric Model for Dark Matter with U(1) Portal}
%
%
%
%
%
%
%

\section{Introduction}
\label{sec:1}

 Left-Right Symmetric Model (LRSM), based on the gauge group $SU(2)_{L} \times SU(2)_{R} \times U(1)_{B-L}$
\cite{PatiPRD1974,MohapatraPRD1975,SenjanoviPRD1975}, was originally proposed to understand the origin of
parity violation in the standard model (SM).  The fact that the seesaw mechanism for understanding small neutrino masses~\cite{seesaw1,seesaw2,seesaw3,seesaw4,seesaw5} finds a natural home in these models, has made them more interesting for theory as well as experiments. In particular, the fact that the $SU(2)_R$ breaking scale is allowed to be in the TeV range by low energy flavor changing neutral current constraints~\cite{zhang,miha} has  provided a strong motivation to look for signatures of neutrino mass related physics in colliders
such as the Large Hadron Collider (LHC) and low energy processes such as neutrinoless double beta-decay.
The current LHC limit on $M_{W_R}$ is 4.4 TeV~\cite{CMS} from a study of the $\ell\ell jj$ final states and
an ATLAS bound of 3 TeV from a study of $tb$ final states~\cite{ATLAS}.
There have also been recent speculations that an anomaly in understanding the CP violating parameter $\epsilon'/\epsilon$
in the SM can possibly be resolved in the left-right models with TeV scale $W_R$~\cite{cirigliano,haba}.

In this paper, we explore an extension of this model to understand the origin of dark matter (DM). Two key questions are (i) whether the DM particle is naturally stable and (ii) how  it is connected to the SM sector. In recent years, an interesting class of models has been proposed where by adding certain fermion or scalar multiplets to the LRSM makes them naturally stable~\cite{heeck,olive}  due to symmetries already present in the model and provide candidates for dark matter. It is then connected to the SM sector via the $W_R$ and $Z_R$ bosons. There have been extensive discussions of DM phenomenology in these models~\cite{heeck2,hooper,patra2,patra3}\footnote{For a warm dark matter possibility in the minimal left-right model, see~\cite{goran}}. The most minimal of these models has triplet fermions with $B-L=0$. The neutral member of this triplet is the dark matter. This triplet model also leads to coupling unification without the need for supersymmetry~\cite{amitabha}, which is an interesting property. The structure of these models however implies that there must be constraints between the $W_R$ mass ($M_{W_R}$) and
the DM mass ($M_{DM}$) for the neutral member of the triplet to be the lightest and hence be a viable dark matter. In terms of particle content, the model has six new fermionic states in addition to the usual particle content of the LRSM.

In this paper, we present a slightly more minimal (in the sense of particle content) alternative extension of the left-right model,
which also provides a dark matter fermion.
The model is based on an extended gauge group $SU(2)_{L} \times SU(2)_{R} \times U(1)_{B-L} \times U(1)_{X}$
with a left-right singlet fermion (called $\zeta_{L,R}$) being a non-singlet under the extra $U(1)_{X}$ as well as under $  U(1)_{B-L}$
in such a way that it is electrically neutral.
The extra $U(1)_X$ provides a gauge portal to the SM fermions with interesting implications.
The detailed phenomenology of dark matter is also very different from the above class of models.

The electric charge formula for this model is
\begin{eqnarray}\label{Qem}
Q_{em}~=~I_{3L}+I_{3R}+\frac{Q_{BL}}{2}+\frac{X}{2} \; .
\end{eqnarray}
The new gauge bosons in the model are $W_{R}$, $Z_{R}$ and the third neutral vector boson,
that we call $X$, will provide the $U(1)$ portal to dark matter in our model.\footnote{For some other examples of models with $U(1)$ portal to DM, see for instance~\cite{farinaldo,MJNeves2018, gu, raut, Okada}.}
In this paper, we will assume that all these bosons are in the multi-TeV mass range.
An interesting implication of this model is that the extra neutral gauge boson, $X$, could be lighter
than the current LHC bounds for the sequential $Z'$ boson \cite{LHCZ1,LHCZ} with mass closer to or even below  a TeV.
We find two parameter ranges for the masses of the DM fermion $\zeta$ and the $X$ boson
  where we can avoid the stringent bounds on the elastic scattering DM cross section with nuclei
  by the direct DM detection experiments, in particular, the XENON1T experiment~\cite{X1T}.

%

The paper is organized as follows:
In Section \ref{sec:2}, we present the model content of
$SU(2)_{L} \times SU(2)_{R} \times U(1)_{B-L} \times U(1)_{X}$ and the Higgs sector. 
Section \ref{sec:3} is dedicated to the sector of gauge bosons to obtain their masses
and the charged and neutral current interactions,
after the spontaneous symmetry breaking (SSB) of the gauge symmetry.
In Section \ref{sec:4}, we calculate the DM relic abundance and identity the model parameter region to reproduce
the observed relic density.
The constraint from the current direct DM detection experiments, in particular, the XENON1T experiment
is derived in Section \ref{sec:5}.
Finally, our concluding comments are cast in Section \ref{sec:6}.

\section{The model content and the Higgs sector}
\label{sec:2}

Our model is based on the gauge group ${\cal G}_{\rm LR}\equiv SU(3)_c\times SU(2)_L\times SU(2)_R\times U(1)_{B-L}\times U(1)_X$.
The quarks and leptons are assigned to the following irreducible representations under ${\cal G}_{\rm LR}$:
\begin{eqnarray}
Q_{L,i} &=&  \left(\begin{array}{c}u_L\\d_L \end{array}\right)_i ,  \left({ \bf 3}, {\bf 2}, {\bf 1}, +\frac{1}{3}\right) \; ,
\nonumber \\
Q_{R,i}  &=&  \left(\begin{array}{c}u_R\\d_R \end{array}\right)_i ,  \left({ \bf 3}, {\bf 1}, {\bf 2}, +\frac{1}{3}\right) \; ,
\nonumber \\
\psi_{L,i}  &=&   \left(\begin{array}{c}\nu_L \\ e_L \end{array}\right)_i , \left({ \bf 1}, {\bf 2}, {\bf 1}, -1 \right) \; ,
\nonumber \\
\psi_{R,i} &=&  \left(\begin{array}{c} N_R \\ e_R \end{array}\right)_i , \left({ \bf 1}, {\bf 1}, {\bf 2}, -1 \right) \; ,
\label{lrSM}
\end{eqnarray}
where $i=1, 2, 3$ is the generation index.
The $U(1)_X$ charges are zero for all these fermions and are not shown in the above equations.
To this fermion content, we add a $SU(2)_L \times SU(2)_R$ singlet vectorlike fermions $\zeta_{L}$ and $\zeta_{R}$
which have $U(1)_{B-L}\times U(1)_X$ charges of $(+a, -a)$. 
In the minimal version of left-right model, the Higgs sector consists of the following multiplets:
\begin{eqnarray}
\Phi \ = \ \left(\begin{array}{cc}\phi^0_1 & \phi^+_2\\\phi^-_1 & \phi^0_2\end{array}\right) : ({\bf 1}, {\bf 2}, {\bf 2}, 0), \qquad \nonumber\\
\Delta_L=\left(\begin{array}{cc}\Delta^+_L/\sqrt{2} & \Delta^{++}_L\\\Delta^0_L & -\Delta^+_L/\sqrt{2}\end{array}\right) : ({\bf 1}, {\bf 3}, {\bf 1}, 2),
\nonumber \\
\Delta_R \ = \ \left(\begin{array}{cc}\Delta^+_R/\sqrt{2} & \Delta^{++}_R\\\Delta^0_R & -\Delta^+_R/\sqrt{2}\end{array}\right) : ({\bf 1}, {\bf 1}, {\bf 3}, 2).
\label{eq:scalar}
\end{eqnarray}
These fields are all chosen to be singlets under the $U(1)_X$ and $SU(3)_c$  gauge groups.
To this Higgs sector, we add a 
Higgs field $\Xi$ with a gauge quantum number  $({\bf 1}, {\bf 1}, b,-b)$
whose vacuum expectation value (VEV) breaks the $U(1)_X$ gauge symmetry.

%
%
%

The Higgs sector of the model and the symmetry breaking are governed by the scalar potential \cite{SenjanovicPRD,kayser,dev,miha0, miha1,goran2,kiers,baren}:
\begin{eqnarray}\label{LHiggs}
V(\Phi, \Delta_{L,R},\Xi) &= & - \mu_1^2 \: {\rm Tr} (\Phi^{\dag} \Phi) - \mu_2^2
\left[ {\rm Tr} (\tilde{\Phi} \Phi^{\dag}) + {\rm Tr} (\tilde{\Phi}^{\dag} \Phi) \right]
- \mu_3^2 \:  {\rm Tr} (\Delta_R
\Delta_R^{\dag}) \nonumber
\\
&&+ \lambda_1 \left[ {\rm Tr} (\Phi^{\dag} \Phi) \right]^2 + \lambda_2 \left\{ \left[
{\rm Tr} (\tilde{\Phi} \Phi^{\dag}) \right]^2 + \left[ {\rm Tr}
(\tilde{\Phi}^{\dag} \Phi) \right]^2 \right\} \nonumber \\
&&+ \lambda_3 \: {\rm Tr} (\tilde{\Phi} \Phi^{\dag}) \, {\rm Tr} (\tilde{\Phi}^{\dag} \Phi) +
\lambda_4 \: {\rm Tr} (\Phi^{\dag} \Phi) \left[ {\rm Tr} (\tilde{\Phi} \Phi^{\dag}) + {\rm Tr}
(\tilde{\Phi}^{\dag} \Phi) \right]  \nonumber\\
&& + \rho_1  \left[ {\rm
Tr} (\Delta_R \Delta_R^{\dag}) \right]^2 
+ \rho_2 \: {\rm Tr} (\Delta_R
\Delta_R) {\rm Tr} (\Delta_R^{\dag} \Delta_R^{\dag}) \nonumber
\\
&&+ \alpha_1 \: {\rm Tr} (\Phi^{\dag} \Phi) {\rm Tr} (\Delta_R \Delta_R^{\dag})
+ \left[  \alpha_2 \, e^{i \delta_2}  {\rm Tr} (\tilde{\Phi}^{\dag} \Phi) {\rm Tr} (\Delta_R
\Delta_R^{\dag}) + {\rm H.c.} \right]
\nonumber \\
&&+ \alpha_3 \: {\rm
Tr}(\Phi^{\dag} \Phi \Delta_R \Delta_R^{\dag})+
\beta_{1} \, \text{Tr}[\Phi\Delta_{R}\Phi^{\dagger}\Delta_{L}^{\dagger}]
+\beta_{2} \, \text{Tr}[\tilde{\Phi}\Delta_{R}\Phi^{\dagger}\Delta_{L}^{\dagger}]
\nonumber \\
%
&&-\mu^{\prime \, 2} \, \Xi^{\dagger} \, \Xi+\lambda^{\prime} \left(\Xi^{\dagger} \Xi\right)^{2}
+\eta_{1} \left(\Xi^{\dagger} \Xi\right) \left[ \mbox{Tr}\left(\Delta_{L}^{\dagger} \Delta_{L}\right)\right]
\nonumber \\
&& +\eta_2 \left(\Xi^{\dagger} \Xi\right)
\left[\mbox{Tr}\left(\Phi^\dagger\Phi\right)\right]
+L \leftrightarrow R \; ,
\end{eqnarray}
where $\mu_{i}$ $(i=1, 2, 3)$,  $\mu^\prime$, $\lambda_{i}$ ($i=1,2,3,4$), $\lambda^\prime$, $\rho_i$ $(i=1,2)$,  $\alpha_{i}$ $(i=1,2, 3)$
  and $\eta_{i}$ $(i=1,2)$ are real parameters.
Parity symmetry implies that the model has three independent gauge coupling constants:
 $g_{L}=g_{R}=g$, $g_{BL}$ and $g_X$.
Note that this Higgs potential is invariant under the parity symmetry.

The Yukawa couplings of the model are:
\begin{eqnarray}
{\cal L}_Y =
&+&h_{ij}^{(\ell)} \, \overline{L}_{iL} \, \Phi \, L_{jR}
+  \widetilde{h}_{ij}^{(\ell)} \, \overline{L}_{iL} \, \tilde{\Phi} \, L_{jR}
+h_{ij}^{(Q)} \, \overline{Q}_{iL} \, \Phi \, Q_{jR}
+ \, \widetilde{h}_{ij}^{(Q)} \, \overline{Q}_{iL} \, \tilde{\Phi} \, Q_{jR}
\nonumber \\
&+& f_{Lij} L^{T}_{iL} \, {\Delta}_{L} \, L_{jL}
+f_{Rij}  L^{T}_{iR} \, {\Delta}_{R} \, L_{jR}
-m_{\zeta} \, \overline{\zeta}_{L} \, \zeta_{R} + {\rm H.c.} \; ,
\end{eqnarray}
where
$h_{ij}^{(\ell)}$, $\widetilde{h}_{ij}^{(\ell)}$, $h_{ij}^{(Q)}$ and $\widetilde{h}_{ij}^{(Q)}$
 $(i, j =1,2,3)$ are complex Yukawa coupling constants that yield the masses for the fermions of the model,
 and  $m_{\zeta}$ is the Dirac mass of the $\zeta$ fermion which is  the dark matter candidate of the model.
%
The model has seesaw mechanisms of I- and II-type for neutrinos as in the usual left-right model.
%
%
%

The gauge symmetry ${\cal G}_{\rm LR}$ is broken down to $SU(3)_c \times U(1)_{em}$
  by the VEVs of the Higgs fields, which are defined as
 $\langle \phi_1^0 \rangle=v_1/\sqrt{2}$,
 $\langle \phi_2^0 \rangle=v_2/\sqrt{2}$,
 $\langle \Delta_{L}^0 \rangle=v_{L}/\sqrt{2}$,
 $\langle \Delta_{R}^0 \rangle=v_{R}/\sqrt{2}$,
 and $\langle \Xi \rangle=u/\sqrt{2}$.
For simplicity, we choose the hierarchy among VEV scales such that $v_{L} \ll
\left( \, v_{1} \, , \, v_{2} \, \right) \ll u \ll v_{R}$.
In order to yield the right scale of the electroweak symmetry breaking,
 we have a relation of $v=\sqrt{v_{1}^{2}+v_{2}^{2} + v_L^2} \simeq \sqrt{v_{1}^{2}+v_{2}^{2}} =246$ GeV
 and parametrized the VEVs as $v_1=v \sin \beta$ and $v_2=v \cos \beta$
 with  a $\beta$-angle.
%
The sequence of the SSB is as follows:
First, the $SU(2)_R \times U(1)_{B-L}$ symmetry is broken by $v_R$ to yield the heavy gauge bosons $W_{R}$ and $Z_{R}$.
Next, the $U(1)_X$ symmetry is broken by $u$ and a mass of $X$ boson is generated.
The electroweak symmetry breaking is completed by $v_1$ and $v_2$.
The next section is dedicated to describe masses of the gauge bosons and their interactions with fermions
  along with phenomenological constraints.

\section{The masses of gauge bosons and the structure of neutral current interactions}
\label{sec:3}
After the SSB, the charged gauge bosons acquire their masses as
\begin{eqnarray}\label{LGaugemassesXB}
{\cal L}_{mass}^{WW_{R}} &=&
\left(
\begin{array}{cc}
A_{L\mu}^{\, +} \quad & A_{R\mu}^{\, +} \\
\end{array}
\right)
\left(
\begin{array}{cc}
A \; & \quad \; B
\\
\\
B \; & \quad \; D \\
\end{array}
\right)
\left(
\begin{array}{c}
A_{L}^{\,\,\, \mu -}
\\
\\
A_{R}^{\,\,\, \mu -} \\
\end{array}
\right)
\; ,
\end{eqnarray}
where $\sqrt{2} \, A_{L(R)\mu}^{\, \, \, \pm}=A_{L(R)\mu}^{1} \mp i \, A_{L(R)\mu}^{2}$,
$A=g^{2}\left(v^{2}+2v_{L}^{2}\right)/4\simeq g^{2}v^{2}/4$,
$B=-g^2v_{1}v_{2}/2\simeq-g^2v^{2}\sin\beta\cos\beta/2$ and
$D=g^{2}\left(v^{2}+2v_{R}^{2}\right)/4\simeq g^{2}v_{R}^{2}/4$.
This mass matrix is diagonalized by a $SO(2)$ transformation:
\begin{eqnarray}\label{transfA0CGY}
A_{L\mu}^{\pm} &=& \cos\varphi \, W_{L\mu}^{\, \pm} + \sin\varphi \, W_{R\mu}^{\, \pm} \; ,
\nonumber \\
A_{R\mu}^{\pm} &=& -\sin\varphi \, W_{L\mu}^{\, \pm} + \cos\varphi \, W_{R\mu}^{\, \pm} \; ,
\end{eqnarray}
where $\varphi$ is a small mixing angle given by $\tan(2\varphi)\simeq-2v^{2}\sin(2\beta)/v_{R}^{2}$.
Thereby, the mass eigenvalues of $W_L$ (which is identified as the SM $W$ boson) and $W_R$ bosons
are calculated to be
\begin{eqnarray}\label{massesWW_{R}}
M_{W_L} \simeq \frac{gv}{2}
\hspace{0.3cm} \mbox{and} \hspace{0.3cm}
M_{W_{R}} \simeq \frac{gv_{R}}{\sqrt{2}}
\; .
\end{eqnarray}
We find 
  the charged current interactions among neutrinos-leptons with $W_L$ and $W_{R}$ of the form:
\begin{eqnarray}
{\cal L}^{int}_{\,\, W_L,W_{R}}=\frac{g}{\sqrt{2}} \; \overline{\nu}_{iL} \, \, \slash{\!\!\!\!W}^{+} \ell_{iL}
+\frac{g}{\sqrt{2}} \; \overline{N}_{iR} \, \, \slash{\!\!\!\!W}_{R}^{\, +} \ell_{iR}+\mbox{H.c.} 
\end{eqnarray}

The neutral gauge boson sector consists of four vector fields: $A_{L}^{\, \mu \, 3}$, $A_{R}^{\, \mu \, 3}$, $B^{\mu}$ and $C^{\mu}$.
After the SSB, the neutral gauge boson mass matrix can be cast in the form:
%
\begin{eqnarray}
{\cal L}_{mass}=
 \frac{1}{2} \, \eta_{\mu\nu} \left(V^{\mu}\right)^{T}  M^{2} \, V^{\nu} ,
\end{eqnarray}
where $(V^{\mu})^{T}=\left( \; A_{L}^{\,\,  \mu \, 3} \; \; A_{R}^{\,\, \mu \, 3} \; \; B^{\mu} \; \; C^{\mu} \; \right)$,
%
%
and the mass matrix $M^{2}$ is given by
\begin{equation}\label{MatrixMassa}
M^{2}=
\left(
\begin{array}{cccc}
v_{L}^{2}g^{2} + \frac{v^2}{4}g^2 & -\frac{v^2}{4}g^{2} & -gg_{BL}v_{L}^2 & 0
\\
\\
-\frac{v^2}{4}g^{2} & v_{R}^{2}g^{2}+\frac{v^2g^{2}}{4} & -gg_{BL}v_{R}^2 & 0
\\
\\
-gg_{BL}v_{L}^2 & -gg_{BL}v_{R}^2 & (v_L^2  + v_{R}^2)g_{BL}^{2}+\frac{u^2g_{BL}^{2}}{9} & -\frac{g_{BL}g_Xu^2}{9}
\\
\\
0 & 0 & -\frac{g_{BL}g_Xu^2}{9} & \frac{g_X^2u^2}{9} \\
\end{array}
\right) \, .
\end{equation}
Here, we have chosen $b=+1/3$, for simplicity.
In diagonalizing the mass matrix, we carry out an $SO(4)$ transformation $V \; \longmapsto \; \tilde{V}= R^{t} \, V $,
  where $R$ is 
  an orthogonal matrix belonging to $SO(4)$.
Once the massless photon mode is taken out, the remaining $3\times 3$ matrix can be diagonalized
  by an $SO(3)$ matrix parameterized by three angles.
%
%
%
The neutral gauge boson mass eigenvalues in the mass eigenstate basis $(Z, Z_R, X, A)$ are given by (upto small mixings that we ignore)
\begin{eqnarray}\label{MatrixMassaZZ_{R}A'}
M_{diag}^{2}=
\left(
\begin{array}{cccc}
M_{Z}^{\, 2} & 0 & 0 & 0 \\
0 & M_{Z_{R}}^{\, 2} & 0 & 0 \\
0 & 0 & M_{X}^{\, 2} & 0 \\
0 & 0 & 0 & 0 \\
\end{array}
\right) \; .
\end{eqnarray}
%
%
%
In terms of symmetry breaking VEVs, the mass eigenvalues are given by
\begin{eqnarray}
M_{Z} &\simeq& \frac{gv}{2} \,
\sqrt{1 + \frac{g_{BL}^{2} g^2_X} {g_{BL}^{2}g^2+g^2_Xg^{2}+g^2_Xg^2_{BL}}} \; ,
\nonumber \\
M_{Z_{R}} &\simeq& v_{R} \, \sqrt{ g^{2}+g_{BL}^2 } \; ,
\nonumber \\
M_{X} &\simeq& \frac{u}{3} \,
\sqrt{g_X^2 + \frac{g^2 \, g_{BL}^{2} }{g^{2}+g_{BL}^{2}}} \; .
\label{massesZZ'}
\end{eqnarray}
The zero mass eigenstate that emerges in Eq.~(\ref{MatrixMassaZZ_{R}A'}) is identified as the electromagnetic massless photon,
   which is expressed as a linear combination of the original gauge bosons:
\begin{eqnarray}
\frac{A^{\mu}}{e}=~\frac{W_{3L}^{\mu}+W_{3R}^{\mu}}{g}+\frac{B^{\mu}}{g_{BL}}+\frac{C^{\mu}}{g_X} \; .
\end{eqnarray}
Defining
\begin{eqnarray}
\frac{e}{g_{BL}}=\sqrt{\cos 2\theta_W}\sin \theta_1 \; ,
\nonumber\\
\frac{e}{g_{X}}=\sqrt{\cos 2\theta_W}\cos \theta_1 \; ,
\end{eqnarray}
%
and using the parity symmetry to set $g_L=g_R=g$,  we obtain the masses of $W_L$ and $Z$
just like in the Glashow-Weinberg-Salam (GWS) model: $M_{W} \simeq 80$ GeV  and $M_{Z} \simeq 91$ GeV.
Here, we have used $v=246$ GeV and the weak mixing angle of $\sin^{2}\theta_{W} \simeq 0.23$,
 and have fixed the fundamental charge by the fine structure constant $e^{2}=4\pi/128$ at the $M_Z$ scale.

Note that with the relation
\begin{eqnarray}\label{gg'}
g_{BL} \sin\theta_{1}=g_{X} \cos\theta_{1}=\frac{e}{\sqrt{\cos(2\theta_{W})}} \simeq 0.43 \; ,
\end{eqnarray}
two parameters, $\theta_{1}$ and $g_X$, are expressed in terms of $g_{BL}$ as
$\sin\theta_{1}=0.43/g_{BL}$ and $g_X= g_{BL} \tan \theta_1= 0.43g_{BL}/\sqrt{g_{BL}^2-(0.43)^2}$.
Thereby, we can use $g_{BL} \geq 0.43$ as a free parameter of the model for our analysis in the following sections.
In this parameterization, we can express Eq.~(\ref{massesZZ'}) as
\begin{eqnarray}
M_{Z} &\simeq& \frac{e \, v }{\sin\left(2\theta_W\right)} \; ,
\nonumber \\
M_{Z_{R}} &\simeq& v_{R} \, \sqrt{ \frac{e^2}{\sin^2 \theta_W} +g_{BL}^2}  \; ,
\nonumber \\
M_{X} &\simeq&
\frac{g_{BL} u}{3} \, \sqrt{ \frac{\left(0.43 \right)^{2}}{g_{BL}^{\, 2}-\left(0.43\right)^{2}}+\frac{e^2}{e^2+g_{BL}^{2}\sin^{2}\theta_{W}} } \; .
\label{massesZZ'-2}
\end{eqnarray}
As a benchmark value, we choose the $W_{R}$  mass at the lower limit obtained by the LHC experiment, $M_{W_{R}}=4.4$ TeV \cite{CMS,ATLAS},
 which means
$v_{R} \simeq 9.6 \, \mbox{TeV}$.
This leads to the $M_{Z_R}$ value as
\begin{eqnarray}
M_{Z_{R}} &\simeq&
\sqrt{
2  M_{W_{R}}^{\, 2} + g_{BL}^2 v_R^2
}
\geq
\sqrt{
2  M_{W_{R}}^{\, 2} + (0.43)^2 \, v_R^2
} \simeq 7.5 \; \rm {TeV}.
\end{eqnarray}
%

%
%
%

The mass eigenstates are approximately expressed in terms of the original fields as
\begin{eqnarray}
W_{L}^{\,\,  \mu \, 3} &=& \cos\theta_{W} Z^{\mu}
+ \sin\theta_{W} A^{\mu} \, ,
\nonumber \\
W_{R}^{\,\, \mu \, 3} &=& - \tan\theta_{W}\sin\theta_{W} Z^{\mu} + \sin\theta_{W} A^{\mu}
+ \frac{\sqrt{\cos(2\theta_{W})}}{\cos\theta_{W}} \, Z_{R}^{\, \mu}
\, ,
\nonumber \\
B^{\mu} &=&
\sin\theta_{1} \sqrt{\cos(2\theta_W)} \left( \, - \tan\theta_{W} Z^{\mu} + A^{\mu} \, \right)
-\sin\theta_{1}\tan\theta_{W} \, Z_{R}^{\, \mu}+\cos\theta_{1} \, X^{\mu}
\, ,
\nonumber \\
C^{\mu} &=& \cos\theta_1 \sqrt{\cos(2\theta_W)} \left( \, - \tan\theta_{W} Z^{\mu} + A^{\mu} \, \right)
-\cos\theta_1 \tan\theta_{W} \, Z_{R}^{\, \mu}-\sin\theta_{1} X^{\mu}
\; .
\; \;
\end{eqnarray}
We ignore small mixings of order $(\frac{v}{v_R}, \frac{u}{v_R}, \frac{v}{u})$ among them in our discussion below.
The neutral bosons, $Z$, $Z_{R}$, $X$ and $A$, interact with a chiral fermion $f_{L(R)}$ (left- or right-handed) of the model as
\begin{equation}\label{LintAZC}
{\cal L}^{int}=-e \, \bar{f}_{L(R)} \left( \, Q_{em} \, \slash{\!\!\!\!A} + Q_{Z} \, \slash{\!\!\!\!Z}+ Q_{Z_{R}} \, \slash{\!\!\!\!Z}_{R}
+ Q_{X} \, \slash{\!\!\!\!X} \, \right) f_{L(R)} \; .
\end{equation}
%
%
%
Here, the charge generators, $Q_{Z}$, $Q_{Z_{R}}$ and $Q_{X}$, are described as
\begin{eqnarray}
Q_{Z} &=&  \frac{I_{L}^{3} -\sin^2\theta_{W} Q_{em}}{\sin\theta_{W}\cos\theta_{W}} \; ,
\nonumber \\
Q_{Z_{R}} &=& \frac{I_{R}^{3} + \tan^{2}\theta_{W} \left(I_{L}^{3}-Q_{em} \right)}{\tan\theta_{W}\sqrt{\cos(2\theta_{W})}} \; ,
\nonumber \\
Q_{X} &=& \frac{\cot\theta_{1} \, Q_{BL}- \tan\theta_{1} \, X}{2\sqrt{\cos(2\theta_{W})}} \; ,
\end{eqnarray}
and the electric charge $Q_{em}$ is given by Eq.~(\ref{Qem}).
All the neutral currents of the model are contained below:
\begin{eqnarray}
{\cal L}_{NC}^{int} &=& e \, J_{\mu}^{em} A^{\mu}+\frac{g}{\cos\theta_{W}} J_{\mu}^{0} \, Z^{\mu}+
\frac{g \cos\theta_{W}}{\sqrt{\cos(2\theta_{W})}} J_{Z_{R} \, \mu}^{0} Z_{R}^{\, \mu}+\frac{g_{X}}{\sin\theta_{1}} J_{X \, \mu}^{0} X^{\mu} \; ,
\end{eqnarray}
where $J_{\mu}^{em}$ is the usual electromagnetic current, $J_{\mu}^{0}=J_{L \, \mu}^{3}-\sin^2\theta_{W} \, J_{\mu}^{em}$
is the $Z$-neutral current of the GSW model, and the neutral currents of $Z_{R}$ and $X$ are given by
\begin{eqnarray}
J_{Z_{R} \, \mu}^{0} &=& J_{R \, \mu}^{3}+\tan^2\theta_{W} \left(J_{L \, \mu}^{3}  - J_{\mu}^{em} \right) \; ,
\nonumber \\
J_{X \, \mu}^{0} &=& J_{\mu Q}-\sin^{2}\theta_{1}\left( J_{\mu}^{em}-J_{L \, \mu}^{3}-J_{R \, \mu}^{3}  \right) \; .
\end{eqnarray}
More explicitly, the neutral currents can be written in terms of a Dirac fermion $f$:
\begin{eqnarray}
J_{Z_{R} \, \mu}^{0} &=& \frac{1}{2} \, g_{L}^{\, f} \, \bar{f} \gamma_{\mu} \left(1-\gamma_{5}\right)f+
\frac{1}{2} \, g_{R}^{\, f} \, \bar{f} \gamma_{\mu} \left(1+\gamma_{5}\right)f \; ,
\nonumber \\
J_{X \, \mu}^{0} &=& \frac{1}{2} \, h_{L}^{\, f} \, \bar{f} \gamma_{\mu} \left(1-\gamma_{5}\right)f+
\frac{1}{2} \, h_{R}^{\, f} \, \bar{f} \gamma_{\mu} \left(1+\gamma_{5}\right)f \; ,
\label{JZ_{R}Ja'}
\end{eqnarray}
%
where $f$ stand for a lepton, a quark or the dark matter particle $\zeta$.
The coefficients $g_{L(R)}^{\, f}$ and $h_{L(R)}^{\, f}$ are defined as
\begin{eqnarray}
g_{L(R)}^{\, f} &=& I_{R}^{\, 3}+\tan^2\theta_{W} \left( I_{L}^{\, 3}- \, Q_{em}^{\, f} \right) \; ,
\nonumber \\
h_{L(R)}^{\, f} &=& Q_{BL}^{f} -\sin^2\theta_{1} \left( Q_{em}^{\, f}-I_{L(R)}^{\, 3} \right) \; .
\end{eqnarray}
%
%
%
%
%
%
%
Here, we list all the expressions of $g^{f}_{L(R)}$ and $h^{f}_{L(R)}$
 in terms of coupling constant $g_{BL}$ and $\theta_{W}$:
\begin{eqnarray}
g^{\ell_{i}}_{L} &=& + \frac{1}{2} \, \tan^2\theta_{W}
\hspace{0.3cm} , \hspace{0.3cm}
g^{\ell_{i}}_{R}=-\frac{1}{2} + \tan^2\theta_{W} \;,
\nonumber \\
g^{\nu_{i}}_{L} &=& + \frac{1}{2} \, \tan^2\theta_{W}
\hspace{0.3cm} , \hspace{0.3cm}
g^{\nu_{i}}_{R}=+\frac{1}{2} \; ,
\nonumber \\
g^{u,c,t}_{L} &=& -\frac{1}{6} \, \tan^2\theta_{W}
\hspace{0.3cm} , \hspace{0.3cm}
g^{u,c,t}_{R}= \frac{1}{2}-\frac{2}{3} \, \tan^2\theta_{W} \; ,
\nonumber \\
g^{d,s,b}_{L} &=& -\frac{1}{6} \, \tan^2\theta_{W}
\hspace{0.3cm} , \hspace{0.3cm}
g^{d,s,b}_{R}= -\frac{1}{2}+\frac{1}{3} \, \tan^2\theta_{W} \; ,
\nonumber \\
g^{\zeta}_{L} &=& 0
\hspace{0.3cm} , \hspace{0.3cm}
g^{\zeta}_{R}= 0 \; ,
\end{eqnarray}
and
\begin{eqnarray}\label{hcoefficients}
h^{\ell_{i}}_{L(R)} &=& h^{\nu_{i}}_{L(R)} = - \frac{1}{2} \, \cos^2 \theta_1 = - \frac{1}{2} \, \left[1-\left(\frac{0.43}{g_{BL}} \right)^2 \right]
\hspace{0.3cm} , \hspace{0.3cm}
\nonumber \\
\nonumber \\
h^{u,c,t}_{L(R)} &=& h^{d,s,b}_{L(R)} = + \frac{1}{6} \, \cos^2 \theta_1=+\frac{1}{6} \, \left[1-\left(\frac{0.43}{g_{BL}} \right)^2 \right]
\hspace{0.3cm} , \hspace{0.3cm}
\nonumber \\
\nonumber \\
h^{\zeta}_{L(R)} &=& +a \; .
\end{eqnarray}
The interaction of neutrinos/leptons with the $Z_{R}$ boson has the
contribution of both vector and axial currents,
\begin{eqnarray}
{\cal L}_{\bar{\nu}_{i} \nu_{i}Z_{R}}^{int}=\frac{g\cos\theta_{W}}{\sqrt{\cos(2\theta_{W})}} \, 
Z_{R \, \mu}\left[\bar{N}_R\gamma^\mu N_R+\tan^2\theta_W\bar{\nu}_L\gamma^\mu \nu_L\right]%
\end{eqnarray}
and
\begin{eqnarray}
{\cal L}_{\bar{\ell}_{i} \ell_{i} Z_{R}}^{int}= \frac{g}{4\cos\theta_{W}\sqrt{\cos(2\theta_{W})}}
\,
\bar{\ell}_{i} \, \, \slash{\!\!\!\!Z}_{R} \! \left[ \phantom{\frac{1}{2}} \!\!\!\! \cos(2\theta_{W})-2\sin^{2}\theta_{W}
+\cos(2\theta_{W}) \, \gamma_{5} \right] \ell_{i} \, .
\end{eqnarray}

To obtain the decay width formula, it is convenient to write the neutral currents of Eq.~(\ref{JZ_{R}Ja'})
in terms of vector and axial-vector components and we find the partial decay with of the $Z_{R}$ boson into a fermion $f$ as
\begin{eqnarray}\label{DecayZ_{R}}
&&
\Gamma(Z_{R} \rightarrow \bar{f} \, f )= \frac{G_{F} M_{Z_{R}} }{24\pi\sqrt{2}}
 \frac{M_{Z}^{2}\cos^4\theta_{W}}{\cos(2\theta_{W})}
\left( |g^{f}_{V}|^2+|g^{f}_{A}|^2 \right)
\sqrt{1-4\, \frac{m_{f}^{\, 2}}{M_{Z_{R}}^{\, 2}}} \left( \, 1
+\, \frac{2m_{f}^{\, 2}}{M_{Z_{R}}^{\, 2}} \, \right) \; ,
\end{eqnarray}
where $G_F=(\sqrt{2} v^2)^{-1}$ is the Fermi constant,
 $m_f$ is the fermion mass satisfying $2 m_f < M_{Z_{R}}$,
and the vector and axial-vector couplings are defined as
$g_{V/A}^{\, f}=2\left(g_{L}^{\, f} \pm g_{R}^{\, f}\right)$, respectively.
For example, the partial decay widths into charged-leptons and neutrinos (with the right-component contribution)
 are given by
\begin{eqnarray}
\Gamma(Z_{R} \rightarrow \bar{\ell}_{i} \, \ell_{i} ) \!&=&\!
\frac{e^{2} M_{Z_{R}}}{192\pi} \, \frac{1-4\tan^2\theta_{W}+5\tan^{4}\theta_{W}}{\sin^2\theta_{W}\cos(2\theta_{W})}    \; ,
\nonumber \\
\Gamma(Z_{R} \rightarrow \bar{\nu}_{i} \, \nu_{i} ) \!&=&\! \frac{e^{2} M_{Z_{R}} }{192\pi} \frac{1+\tan^{4}\theta_{W}}{\sin^2\theta_{W}\cos(2\theta_{W})} \; .
\end{eqnarray}
For $M_{Z}=91$ GeV and $M_{Z_{R}}=7.5$ TeV, we obtain 
\begin{equation}
\Gamma(Z_{R} \rightarrow \bar{\ell}_{i} \, \ell_{i} ) \simeq 2.6 \, \mbox{GeV}
\hspace{0.2cm} \mbox{and} \hspace{0.2cm}
\Gamma(Z_{R} \rightarrow \bar{\nu}_{i} \, \nu_{i} ) \simeq 9.8 \, \mbox{GeV} \; .
\end{equation}
%
%
%
%

%
%
%
Using Eqs.~(\ref{JZ_{R}Ja'}) and  (\ref{hcoefficients}), let us express the interaction
 of $X$ boson with a SM fermion $f$ and with the DM fermion $\zeta$ as
\begin{eqnarray}
{\cal L}_{X}^{\, int}=g_{f} \, \bar{f} \, \,  \slash{\!\!\!\!X} \, f + g_{\zeta} \, \bar{\zeta} \, \,  \slash{\!\!\!\!X} \, \zeta \; ,
\end{eqnarray}
where $g_{f}=Q_{BL}^f \, \tilde{g}_{BL}$ with $\tilde{g}_{BL}=\sqrt{1-\left(0.43/g_{BL}\right)^{2}} \left(g_{BL}/2\right)$
  and $Q_{BL}^f$ being the $B-L$ charge of the fermion, and  $g_{\zeta}=2 \, a \, \tilde{g}_{BL}$.
%
The effective coupling $\tilde{g}_{BL}$ is a monotonically increasing function of $g_{BL} \geq 0.43$
  with limits of $\tilde{g}_{BL} \to 0$ for $g_{BL} \to 0.43$ and
  $\tilde{g}_{BL} \simeq g_{BL}$ for $g_{BL}^{2} \gg \left(0.43\right)^{2}$.
Note that even though the electric charge formula implies a lower bound on $g_{BL}$, the effective $X$ coupling $\tilde{g}_{BL}$ can be smaller.
In the following analysis, we use $\tilde{g}_{BL}$ as a free parameter, instead of $g_{BL}$.
Note that the interaction of the $X$ boson with the SM fermions is exactly the same as
  that of the $B-L$ gauge boson ($Z^\prime$ boson) in the minimal $B-L$ model \cite{BL1, BL3, BL4, BL5}
  when identifying $\tilde{g}_{BL}$ with the $B-L$ gauge coupling.
Since the $\zeta$-charge $a$ is a free parameter, we can use $g_{\zeta}$ as a free parameter in our analysis on DM physics.

\begin{figure}[t]
 \vspace{-5pt}
 \centering
  \includegraphics[width=0.7\textwidth]{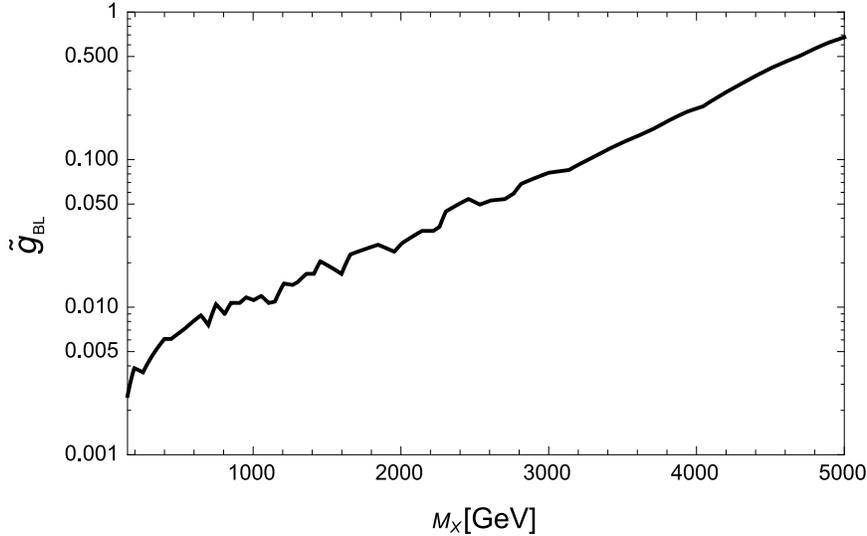} \hspace{0.1cm}
   \caption{
   The ATLAS bound on the effective coupling $\tilde{g}_{BL}$ as a function of $M_X$
   from Figure 5 in Ref.~\cite{LHCZ1}.
   }
 \label{ATLAS}
\end{figure}

The ATLAS and CMS collaborations have been searching for a narrow resonance with the dilepton final states ($e^+ e^-$ and $\mu^+ \mu^-$)
  at the LHC.
As a benchmark model, the production of the $Z^\prime$ boson of the minimal $B-L$ model has been analyzed
 by the ATLAS collaboration with a 36.1/fb luminosity and a collider energy of $\sqrt{s}=13$ TeV at the LHC Run-2 \cite{LHCZ1},
 and the upper bound of  the $B-L$ gauge coupling as a function of $Z^\prime$ boson mass
 has been obtained.\footnote{See, for example, Ref.~\cite{Okada} about how to interpret the upper bound
 on the production cross section of the dilepton final states into the relation between the $B-L$ gauge coupling and
 the $Z^\prime$ boson mass. }
By identifying the $B-L$ gauge coupling and the $Z^\prime$ boson mass with $\tilde{g}_{BL}$ and $M_X$, respectively,
 we show the current ATLAS bound in Figure \ref{ATLAS}.

In the ATLAS analysis, the $Z^\prime$ boson is assumed to decay into only the SM fermions.
If the $X$ boson has additional decay modes into new particles, the upper bound must be modified.
In our model, the $X$ boson can also decay into a pair of DM particles for $2 m_\zeta < M_X$.
As we can see in the next section, the total decay width of the $X$ boson is very small compared to the $X$ mass,
 so that the narrow decay width approximation can be justified to evaluate the $X$ production cross section
 at the LHC Run-2.
In the approximation, the cross section of the process $q {\bar q} \to X \to \ell^+ \ell^-$ at the parton level
  is described as
\begin{eqnarray}
   \sigma(q {\bar q} \to X \to \ell^+ \ell^-) \propto (\tilde{g}_{BL} )^2 \times {\rm BR}(X \to \ell^+ \ell^-) \; .
 \label{narrowDW}
\end{eqnarray}
Hence, if the $X$ boson can decay into a pair of the DM particles, ${\rm BR}(X \to \ell^+ \ell^-)$ becomes
  smaller and as a result, the upper bound on $\tilde{g}_{BL}$ is increasing.
In the presence of the decay of $X \to \bar{\zeta} \zeta$, we scale the result shown in Figure \ref{ATLAS}
  by a factor of
\begin{eqnarray}
   \sqrt{1+\frac{\Gamma(X \to \bar{\zeta} \zeta)}{\Gamma_{SM}}} \; ,
 \label{scaling}
\end{eqnarray}
 where $\Gamma_{SM}$ is the partial decay width of $X$ into all the kinematically allowed SM fermions,
 and $\Gamma(X \to \bar{\zeta} \zeta)$ is the partial decay width into a pair of the DM particles.

%
%
%

%

\section{The Dark matter relic density}
\label{sec:4}

The DM particle in our model is the $\zeta$ fermion which is stable by its representation
  under the gauge group of the model.
Provided $2a\neq b$, $\zeta$ is a Dirac fermion.
Its pair annihilates to SM fermion pairs and a pair of $X$ bosons (if kinematically allowed)
  via its coupling to the $X$ boson.
We assume that $\zeta$ and $X$ are in thermal equilibrium in very early universe.
This already imposes constraints on the parameters of the model.
To see these constraints, note that the conditions for thermal equilibrium between $\zeta$,  SM fermions $f$ and $X$
 in the early universe $\left( T \gg m_{\zeta} , M_{X} \right)$ are
\begin{enumerate}
\item $n_{\zeta}  \sigma(\zeta \, \bar{\zeta} \to f \, \bar{f}) > H$ .  
\item $n_{X}  \sigma(X \, X \to f \, \bar{f}) > H$ .
\end{enumerate}
Here, $H\sim T^2/M_{Pl}$ is the Hubble parameter with the Planck mass, $M_{Pl}=1.22 \times 10^{19}$ GeV.
Using $n_{\zeta,X} \sim T^{3}$ for the number density of $\zeta$ or $X$
 and $\sigma(\zeta\, \bar{\zeta} \to f \, \bar{f}) \sim \frac{g_{\zeta}^{\,2} \, g_{f}^{\, 2}}{T^{2}}$, we obtain
 for the first case
$n_\zeta \sigma (\zeta \, \bar{\zeta} \to f \, \bar{f})=g_{\zeta}^{\,2} \, g_{f}^{\, 2} \, T$.
Similarly, using $\sigma(X \, X \to f \, \bar{f}) \sim \frac{g_{f}^{\, 4}}{T^{2}} $, we obtain
 $n_X \sigma(X \, X \to f \, \bar{f})=g_{f}^{\, 4} \, T $
for the second case.
We define the freeze-in temperature $(T_{FI})$ as
\begin{eqnarray}
\left. n \sigma \right|_{T=T_{FI}}=H(T_{FI})\sim\frac{T_{FI}^{\; 2}}{M_{Pl}} \; ,
\end{eqnarray}
at which $\zeta$ or $X$ gets in thermal equilibrium with the SM particles.
Requiring $T_{FI} >m_{\zeta},M_{X}$, we find the lower bounds,
%
%
$g_{\zeta}^{\, 2} g_{f}^{\, 2}>m_{\zeta}/M_{Pl}$ and $g_{f}^{\, 4}>M_{X}/M_{Pl}$,
for the first and second cases, respectively.
Roughly, $g_{\zeta} \sim g_{f} > 10^{-4}$ satisfies these conditions for $m_\zeta, M_X \sim 1$ TeV.

We are now ready to calculate the relic density of the $\zeta$ fermion.
%
There are two typical annihilation processes: (1) $\zeta\bar\zeta \to X \to f\bar{f}$ though the exchange of $X$ boson in the $s$-channel,
  where $f$'s are the SM fermions.
(2) $\zeta\bar\zeta\to XX$, which is active only for $m_\zeta > M_X$.
In some parameter region, the process (1) dominates the annihilation cross section (Case (i)),
 while the process (2) can dominate over the first process in some parameter region (Case (ii)).
We discuss these two cases below.
As we will see later on in the next section, the direct DM detection constraint is easier to satisfy
  in Case (ii) compared to Case (i) where it can be satisfied only in a narrow range for the masses and couplings.

We first consider Case (i), where the annihilation process is dominated by $\zeta \bar{\zeta} \to X \to f\bar{f}$.
In the non-relativistic limit, the annihilation cross section times relative velocity for this process is given by \cite{farinaldo}
\begin{eqnarray}\label{vsigma}
v_{rel} \sigma(\zeta\bar\zeta\to X \to f\bar{f}) \simeq \frac{g^2_{\zeta}}{2\pi} \sum_{f} N_{c}^{\, f} \, g^2_{f} \,
\frac{2m_{\zeta}^{\, 2}+m^2_f}{(M^2_X-4m^2_\zeta)^2+M^2_X\Gamma^2_X} \, \sqrt{1-\frac{m_f^2}{m_\zeta^2}} \; ,
\end{eqnarray}
where $f$ denotes a SM fermion with mass of $m_f$, and $N_{c}^f$ is the color number in the final state of a SM fermion:
$N_{c}^{f}=3$ for a quark, $N_{c}^{f}=1$ for a charged lepton, $N_{c}^{f}=1/2$ for a SM neutrino $\left(m_{f} \rightarrow 0\right)$.
The total decay width of $X$ boson is given by
\begin{eqnarray}\label{decayX}
\Gamma_X=\frac{M_{X}}{12\pi} \sum_{f} N_{c}^{f} g_{f}^{\, 2} \left(1+\frac{2m_{f}^{\, 2}}{M_{X}^{\, 2}} \right) \sqrt{1-\frac{4m_{f}^{\, 2}}{M_{X}^{\, 2}}} \; ,
\end{eqnarray}
with the condition $M_{X}>2m_{f}$. If it is kinetically allowed, the partial decay width to $X \to \bar{\zeta} \, \zeta$
\begin{eqnarray}\label{decayXzeta}
\Gamma (X \to \bar{\zeta} \, \zeta)=\frac{M_{X}}{12\pi} \, g_{\zeta}^{\, 2} \left(1+\frac{2m_{\zeta}^{\, 2}}{M_{X}^{\, 2}} \right) \sqrt{1-\frac{4m_{\zeta}^{\, 2}}{M_{X}^{\, 2}}}
\end{eqnarray}
 must be added to the total decay width.
Here the important point is that the annihilation cross section is proportional to the product
 $g^2_{\zeta} g^2_{f}= g^2_{\zeta}(Q_{BL}^f \tilde{g}_{BL})^2$.

\begin{figure}[tb]
 \vspace{-5pt}
 \centering
  \includegraphics[width=0.8\textwidth]{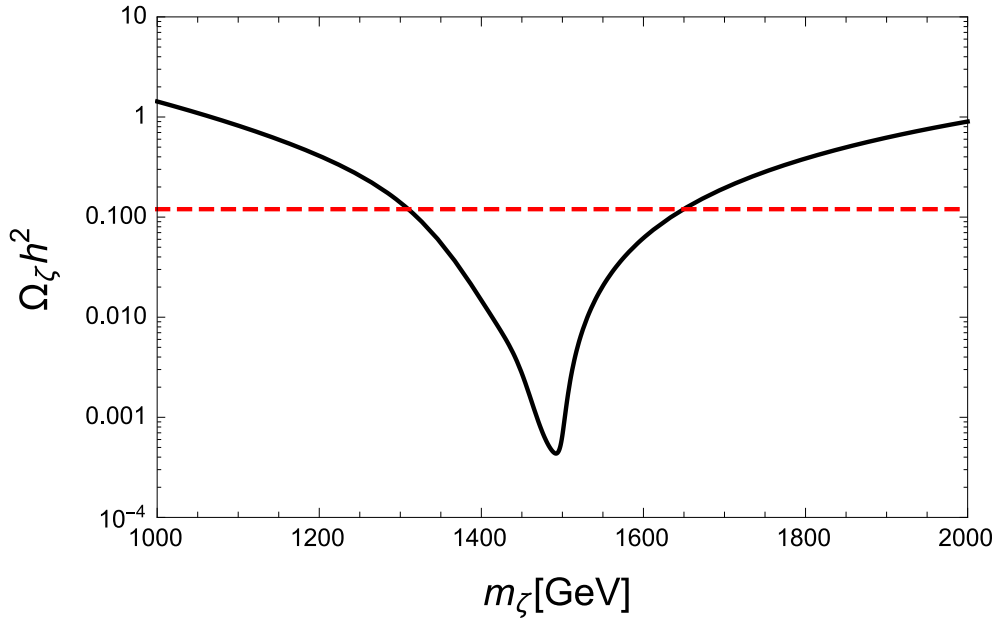} \hspace{0.1cm}
   \caption{
A sample plot of DM relic density vs. $m_{\zeta}$  for $g_{\zeta} = \tilde{g}_{BL}=0.2$ and $M_{X}= 3$ TeV,
 along with the observed relic density (dashed horizontal line).
  }
 \label{DMRD_sample}
\end{figure}

\begin{figure}[t]
 \centering
 \includegraphics[width=0.48\textwidth]{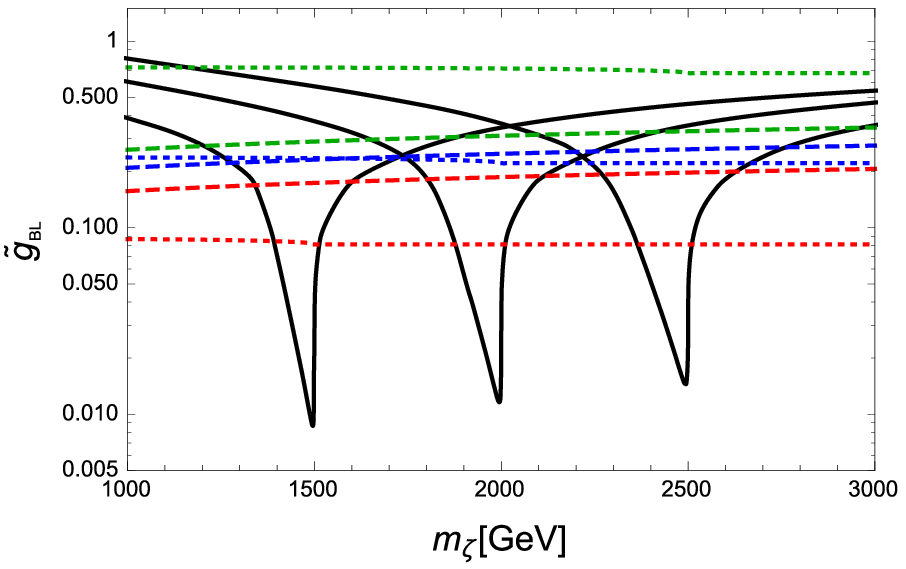} 
 \includegraphics[width=0.48\textwidth]{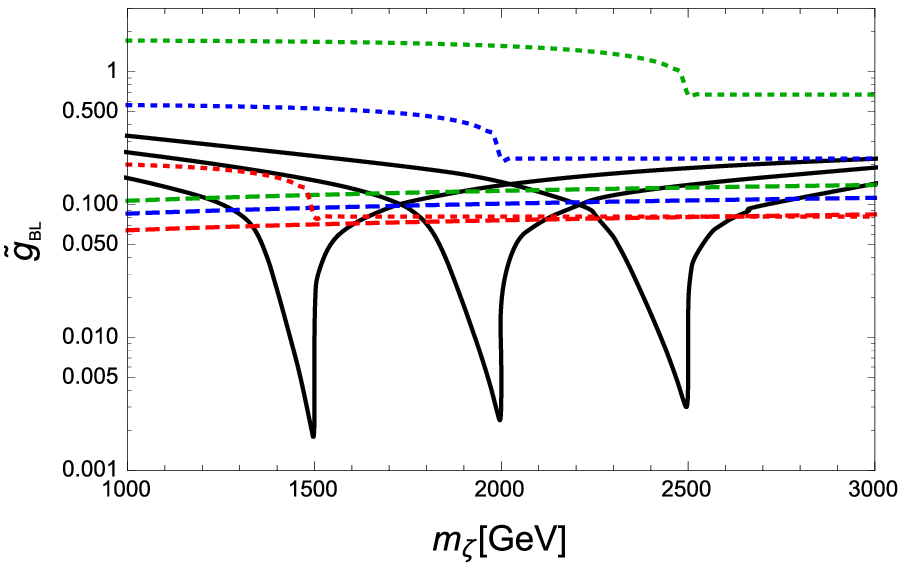}
\caption{
Left panel: Plots of $\tilde{g}_{BL}$ vs $m_{\zeta}$ (solid black lines)
 that give the observed relic density for $M_{X}= 3$, $4$ and $5$ TeV, respectively, from left to right.
Here, we have fixed $g_\zeta=\tilde{g}_{BL}$.
The dotted lines correspond to the ATLAS bounds and dashed lines are the direct detection bounds.
The red, blue and green dashed and dotted lines correspond to the bounds for
  $M_X=3$ , $4$ and  $5$ TeV for from bottom to top.
The dips in the solid lines are where the $M_X\simeq 2 m_\zeta$ and we see that constraints
  from the ATLAS searches for the $Z^\prime$ boson resonance and the direct DM searches are so stringent
  that only the parameter region near the resonance are allowed for Case (i).
Right panel: Same as the left panel but for $g_\zeta= 6 \, \tilde{g}_{BL}$.
}
 \label{DMRD}
\end{figure}

\begin{figure}[t]
 \centering
 \includegraphics[width=0.48\textwidth]{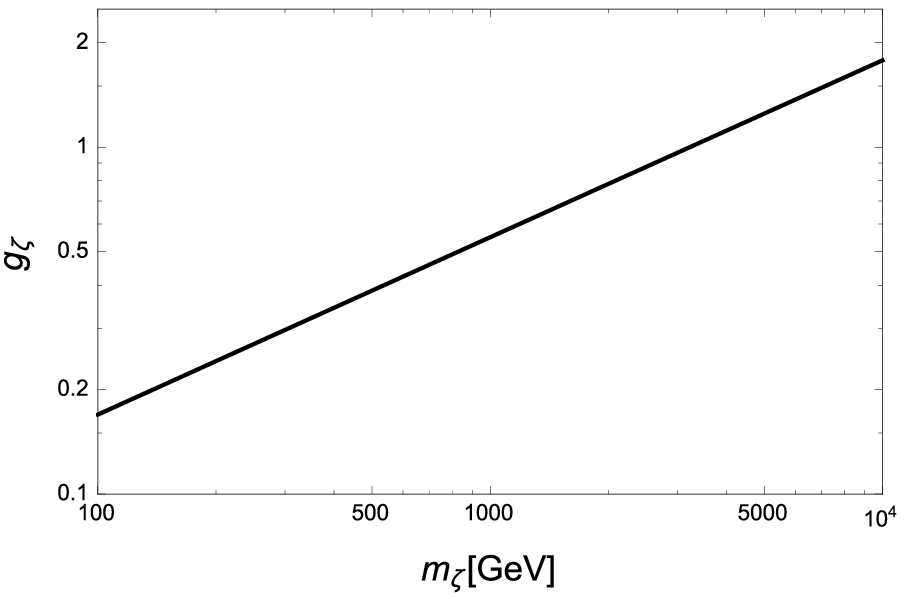} 
 \includegraphics[width=0.5\textwidth]{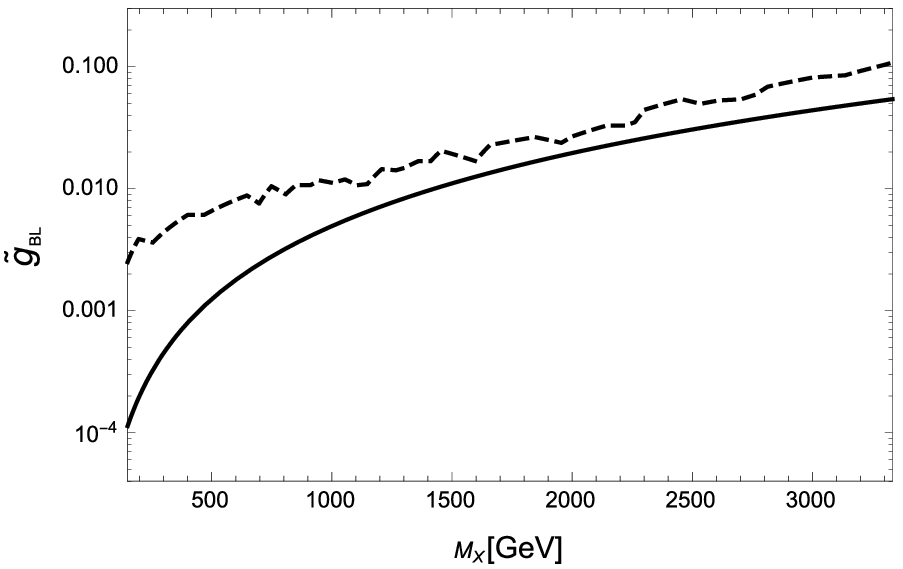}
\caption{
Left panel: Plots of $g_\zeta$ vs $m_{\zeta}$
  along which the observed DM relic abundance is reproduced.
Here, we have set $M_X=m_\zeta/3$.
Right panel: Upper bounds on $\tilde{g}_{BL}$ as a function of $M_X$
  along the solid line in the left panel.
The solid and dashed lines correspond to the XENON1T and the ATLAS bounds, respectively.
}
 \label{XX}
\end{figure}

With a good accuracy, the DM relic density is given as the asymptotic solution of the Boltzmann equation to be~\cite{review1, review2}
\begin{eqnarray}
\Omega_{DM} \, h^{2}
\simeq \frac{ 1.07 \times 10^{9} \, x_{f}}{\sqrt{g_{\ast}} \, M_{Pl} \, \langle v_{rel}\sigma \rangle} \; ,
 \label{Omega}
\end{eqnarray}
where $\langle v_{rel}\sigma \rangle$ is the thermally averaged cross section, and
 the freeze-out temperature of the DM particle is approximately evaluated as
 $x_{f}=m_{DM}/T_{f}\simeq \ln(x)-0.5\ln(\ln(x))$
 with $x=0.038 \sqrt{g_{DM}/g_{\ast}} M_{Pl} \, m_{DM} \, \langle v_{rel}  \sigma \rangle$.
Here, $g_{\ast}$ is the degree of freedom of relativistic particles ($g_{\ast}=106.75$ for the SM particle content), and
 $g_{DM}=4$ is the number of internal degree of freedom for the $\zeta$ fermion.
%
%
Since the annihilation process occurs through $s$-wave, we approximate the thermal averaged cross section
 as $\langle v_{rel} \sigma \rangle \simeq v_{rel} \sigma$.\footnote{
Although Eq.~(\ref{Omega}) is a good approximation,
  in our actual analysis we have numerically solved the Boltzmann equation
  with the thermal averaged cross section which is also numerically evaluated.
Solid black lines in Figures \ref{DMRD_sample}, \ref{DMRD}, \ref{XX} and \ref{Future} show our numerical results.
}

A sample plot of the resultant DM relic density $\Omega_{DM}h^2$ is shown in the Figure \ref{DMRD_sample}
  for $g_\zeta = \tilde{g}_{BL}=0.2$ and $M_X=3$ TeV,
  along with the observed DM relic abundance $\Omega_{DM}h^2 \simeq 0.12$ \cite{Planck2018}.
In this sample plot, $m_\zeta \simeq 1.34$ TeV and $1.68$ TeV reproduce the observed relic abundance.
In Figure \ref{DMRD}, we show the contours on the ($m_\zeta$, $\tilde{g}_{BL}$)-plane
  for $M_X=3$, $4$ and $5$ TeV (solid black lines from left to right) along
  which the observed DM relic abundance is reproduced.
In the left panel, we have fixed $g_\zeta=\tilde{g}_{B-L}$.
In the figure, we have also shown the direct DM detection constraints (see the next section for details)
  and the ATLAS bounds from Figure \ref{ATLAS} with a scaling given in Eq.~(\ref{scaling}).
Due to strong direct detection constraints and the ATLAS bounds,
 the only viable region for DM masses where this case works
 is when we near the resonance point i.e. in the vicinity of $M_X\simeq 2m_\zeta$.
The right-panel shows the same as the left panel, but for the coupling choice of $g_\zeta=6 \, \tilde{g}_{B-L}$.
Since the process $X \to \bar{\zeta} \zeta$ dominates the $X$ boson decay modes,
  the ATLAS bounds for $2 m_\zeta < M_X$  appear weaker than the results shown in the left panel.

Next, we consider Case (ii) where the annihilation process is dominated by  $\bar{\zeta} \, \zeta \to X X$ for $m_{\zeta} > M_{X}$.
In the non-relativistic limit, the annihilation cross section times relative velocity is in this case is given by
\begin{eqnarray}
v_{rel}\sigma(\bar{\zeta} \, \zeta \to X X) \simeq \frac{g_{\zeta}^{\, 4}}{16\pi m_{\zeta}^{\, 2}} \left(1-\frac{M_{X}^{2}}{m_{\zeta}^{\, 2}} \right)^{3/2}
\left(1-\frac{M_{X}^{2}}{2 m_{\zeta}^{\, 2}} \right)^{-2}.
\end{eqnarray}
%
%
Note that the annihilation cross section in this case is independent of $\tilde{g}_{BL}$ and
   this is a crucial difference from Case (i).
By taking $\tilde{g}_{BL}$ as small as possible, we can easily avoid
  the severe constraints from the direct DM detection experiments and the search for $X$ boson at the LHC.

As in Case (i), we employ Eq.~(\ref{Omega}) and evaluate the DM relic density.
In the left panel of Figure \ref{XX}, we show $g_\zeta$ vs.~$m_\zeta$ along which the observed DM relic abundance is reproduced.
In this analysis, we have taken $M_X = m_\zeta/3$ as an example.
We have also calculated the upper bound on $\tilde{g}_{BL}$ as a function of $M_X (=m_\zeta/3)$
  from the ATLAS bounds and the direct DM detection experiments (see the next section).
These upper bounds are shown in the right panel of Figure \ref{XX}.

%

\section{Direct detection of dark matter}
\label{sec:5}

There are constraints on the DM model parameters from the direct detection limits from various experiments, the latest and most stringent one being from the XENON1T experiment~\cite{X1T}.
The key equation is the spin-independent cross section for the elastic scattering of the DM particle with a nucleon $\zeta N \to \zeta N$
  in the model which occurs via the exchange of $X$ boson.
This spin-independent cross section is given by~\cite{farinaldo}
\begin{eqnarray}
\sigma_{SI}\simeq \frac{\mu^2_{\zeta N}}{\pi}\left[\frac{Zf_p+(A-Z)f_n}{A}\right]^{2} \; ,
\end{eqnarray}
where $\mu_{\zeta N}=m_{\zeta}m_{N}/(m_{\zeta}+m_{N})$ is the reduced mass, $m_{N}$ is the nucleon mass, and
\begin{eqnarray}
f_p=\frac{g_{\zeta}}{M^2_X} \left(2g_{u}+g_{d}\right) \; ,
\nonumber\\
f_n=\frac{g_{\zeta}}{M^2_X} \left(g_{u}+2g_{d}\right) \; .
\end{eqnarray}
In our case, $g_{u}=g_{d}=(1/3) \tilde{g}_{BL}$, so that $f_{p}=f_{n}$ and we obtain
\begin{eqnarray}
\sigma_{SI}=\frac{\mu^2_{\zeta N}}{\pi} \, f_{p}^{\, 2}=\frac{1}{\pi} \, g_{\zeta}^{2} \, \tilde{g}_{BL}^{2} \, \frac{\mu_{\zeta N}^{\, 2}}{M_{X}^{\, 4}} \; .
\label{SI}
\end{eqnarray}
From  the XENON1T result \cite{X1T}, we parametrize the upper bound on the nucleon scattering cross section as
\begin{eqnarray}
  \sigma_{SI} \leq 9.0 \times 10^{-11} \, \mbox{pb} \times \left(\frac{m_\zeta}{100 \; {\rm GeV}} \right).
\label{X1T}
\end{eqnarray}
Using Eqs.~(\ref{SI}) and (\ref{X1T}), we have obtained the results shown in Figure \ref{DMRD} (dashed lines)
  and the right panel of Figure \ref{XX} (solid line).

%
%

\section{Conclusions}
\label{sec:6}

We have proposed a minimal extension of the Left-Right Symmetric Model (LRSM)
  for neutrino masses by introducing an extra $U(1)_{X}$ gauge group and a heavy gauge singlet Dirac fermion
  to provide a unified framework for neutrino masses as well as dark matter.
The extra $U(1)$ does contribute to the electric charge formula. The model has an extra neutral
gauge boson, $X$, in addition  to the gauge bosons $W^{\pm}$, $Z$, $W_{R}$ and $Z_{R}$, which plays a key role in the properties of the dark matter. It also connects the dark sector to the visible SM sector. We discuss the constraints on the mass and coupling of this extra gauge boson ($X$), by diagonalizing the $4\times 4$ neutral gauge boson mass matrix to give the approximate eigenstates. 
We find that, depending on the relative mass hierarchy between the $X$-boson and the DM fermion, the allowed parameter space of the DM mass and the DM and SM fermion couplings to $X$ lie in different ranges. The main constraints come from the direct detection bounds as well as the LHC bounds on $X$ production.
These constraints can be easily avoided when a pair of dark matter particles dominantly annihilates to a pair of $X$ bosons.

\begin{figure}[t]
 \centering
\includegraphics[width=0.48\textwidth]{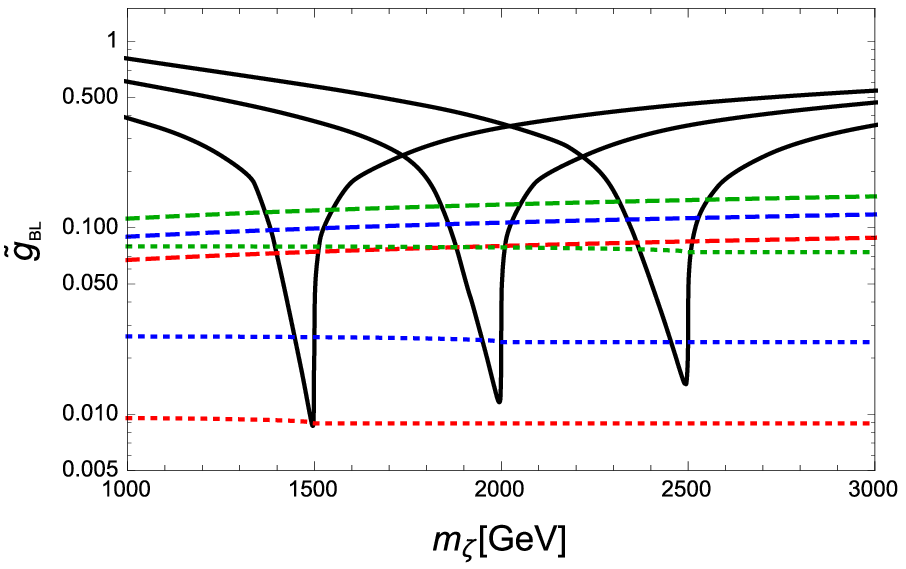}
\includegraphics[width=0.5\textwidth]{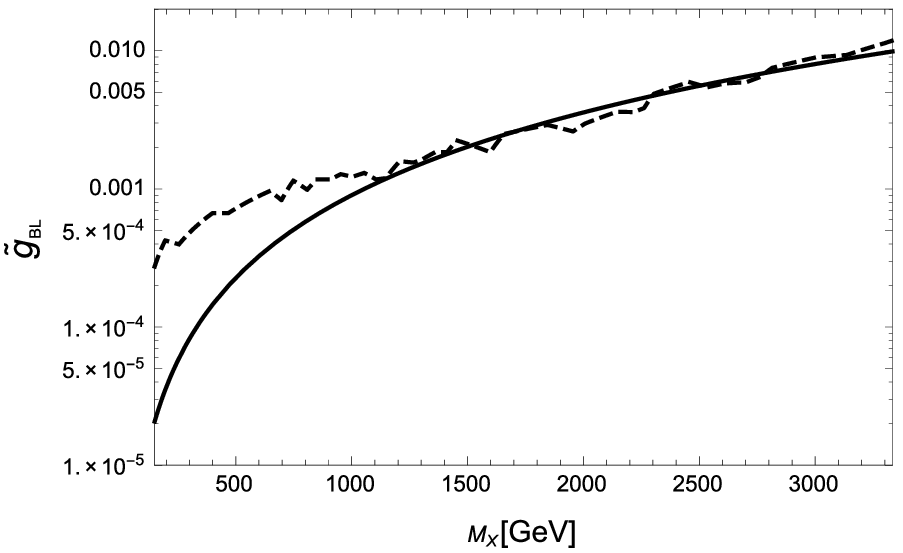} 
\caption{
Left panel: Same as the left panel in Figure \ref{DMRD} but
  with the future reach by the High-Luminosity LHC with a 3000/fb luminosity and the LUX-ZEPLIN DM experiment.
Right panel: Same as the right panel in Figure \ref{XX} but for the future reach.
}
 \label{Future}
\end{figure}

Finally we discuss the prospective bounds on the parameters space
  from the future experiments.
The LUX-ZEPLIN DM experiment \cite{LZ} for the direct DM search is expected to improve
  the current upper bound on the nucleon scattering cross section about one order of magnitude:
\begin{eqnarray}
  \sigma_{SI} \lesssim 3 \times 10^{-12} \, \mbox{pb} \times \left(\frac{m_\zeta}{100 \; {\rm GeV}} \right).
\label{LZ}
\end{eqnarray}
The ATLAS and the CMS collaborations will continue the search for a narrow resonance
  at the LHC with a luminosity upgrade (High-Luminosity LHC).
Since the number of SM background events is very small for a high mass resonance region,
  we naively scale the current upper bound on the cross section with the 36.1/fb luminosity
  to a future bound by a factor of $36.1/3000$ for a 3000/fb integrated luminosity at the High-Luminosity LHC.
Using the narrow decay width approximation (see Eq.~(\ref{narrowDW})),
  we scale the current upper bound on $\tilde{g}_{BL}$ by a factor of $\sqrt{36.1/3000}$.
In Figure \ref{Future}, we show our results with these future prospective bounds.

\section*{Acknowledgement}
The work of R.N.M.~is supported by the US National Science Foundation under Grant No.~PHY1620074
and that of N.O.~is supported by the US Department of Energy under Grant No.~DE-SC0012447.

%


%
\end{document}